\documentclass{aa}  
\usepackage{graphicx}
\usepackage{txfonts}
\begin{document}
\title{Collective pulsational velocity broadening due to gravity modes as a
physical explanation for macroturbulence in hot massive stars}

\author{
C. Aerts\inst{1,2}
\and
J.\ Puls\inst{3}
\and
M.\ Godart\inst{4}
\and 
M.-A.\  Dupret\inst{4}
}
\titlerunning{Macroturbulence in hot massive stars } 
\authorrunning{Aerts, Puls, Godart, Dupret}

\institute{ Instituut voor Sterrenkunde, Katholieke Universiteit Leuven,
Celestijnenlaan 200D, B-3001 Leuven, Belgium\\email: conny@ster.kuleuven.be 
\and
IMAPP, Department of Astrophysics, Radboud University 
Nijmegen, PO Box 9010, 6500 GL
Nijmegen, the Netherlands 
\and 
Universit\"ats-Sternwarte, Scheinerstrasse 1,
D-81679 M\"unchen, Germany
\and
Institut d'Astrophysique et G\'eophysique,
Universit\'e de Li\`ege, all\'ee du Six Ao\^ut 17, B-4000 Li\`ege, Belgium}

\date{Received June 26, 2008; accepted ??}

 
  \abstract{}{We aimed at finding a physical explanation for the
  occurrence of macroturbulence in the atmospheres of hot massive stars, a
  phenomenon found in observations since more than a decade but yet
  unexplained.}  {We computed time series of line
  profiles for evolved massive stars broadened by rotation and by hundreds of
  low-amplitude nonradial gravity-mode pulsations which are predicted to be
  excited for evolved massive stars.}  {In
  general, line profiles based on macrotubulent broadening can mimic those
  subject to pulsational broadening. In several cases, though, good fits require
  macroturbulent velocities that pass the speed of sound for realistic pulsation
  amplitudes. Moreover, we find that the rotation velocity can be seriously
  underestimated by using a simple parameter description for macroturbulence
  rather than an appropriate pulsational model description to fit the line
  profiles.  } {We
  conclude that macroturbulence is a likely signature of the collective effect
  of pulsations.  We provide line diagnostics and their typical values to decide
  whether or not pulsational broadening is present in observed line profiles, as
  well as a procedure to avoid an inaccurate estimation of the rotation
  velocity.}


   \maketitle

%

\section{Introduction of the phenomenon of macroturbulence}

Stars are gaseous bodies that transfer hydrogen into helium through nuclear
burning in their core during the largest part of their lives. A variety of
evolved stars results after the exhaustion of the core hydrogen burning.  It is
the birth mass of the star that determines which evolutionary path the evolved
star will follow. Here, we are concerned with stars whose birth mass is above
ten solar masses. Such massive stars perform subsequent nuclear burning
cycles until their core is composed of iron, after which they collapse as
supernova.  While this broad picture of stellar evolution is well understood and
in agreement with various types of observations, we still lack knowledge of
important aspects of the physics and dynamics inside massive stars and of their
consequences for the stellar life.

One particular shortcoming in the description of the physics of stellar
atmospheres of massive stars is the need to introduce an ad-hoc velocity field,
termed macroturbulence, at the stellar surface in order to bring the observed
shape of spectral lines into agreement with observations. While evidence for the
occurrence of such macroturbulence in hot stars was established since more than
a decade (Howarth et al.\ 1997), there is still no physical explanation for this
phenomenon available. This unsatisfactory situation has become ever more
problematic as the data improved in quality in terms of resolving power and
signal-to-noise (S/N) ratio and in quantity in terms of the number of stars that
have been studied with high-resolution spectroscopy.  It turns out that the
macroturbulent velocities required to explain high-quality observations are
supersonic in many of the studied stars, which would point to highly dynamical
atmospheric motion whose cause is unknown (Ryans et al.\ 2002, Lefever et al.\
2007, Markova \& Puls 2008).  Here, we provide a natural physical explanation
for this phenomenon in terms of the collective effect of numerous stellar
pulsations of low amplitude.

Velocity fields of very different scales occur in the atmospheres of
stars. Apart from the rotational velocity which can vary from zero speed up to
the critical value, line synthesis codes also include a certain amount of {\it
microturbulence\/} (of order a few km\,s$^{-1}$) to bring the observed profiles
in the spectra of stars into agreement with the data. Microturbulence is defined
as a phenomenon related to velocity fields with scales shorter than the mean
free path of the photons in the atmosphere (e.g., Gray 2005 for a thorough
explanation). Microturbulence and rotation are usually treated as
time-independent processes leading to line profile broadening.

In contrast to microturbulence, {\it macroturbulence\/} refers to
velocity fields with a scale larger than the mean free path of the photons (with
mesoturbulence as the intermediate situation -- e.g., Gray 1978). 
Macroturbulence was mainly introduced and studied in the context of 
cool stars (e.g, Gray 1973, 1975, 1978). Various descriptions have been 
proposed in the literature (see Gray 2005 for an overview), among which an
isotropic model and a radial-tangential model are the most common ones.
Both these models will be considered here.

Values for the micro- and macroturbulence are usually derived from line-profile
fits of single snapshot spectra. Here we are focused on such applications to
massive hot stars, whose microturbulent velocities are usually below
15\,km\,s$^{-1}$ (e.g., McErlean et al.\ 1998; Villamariz \& Herrero 2000). The
published values of macroturbulence, on the other hand, are usually well above
this value, reaching up to 90\,km\,s$^{-1}$ (Lefever et al.\ 2007; Markova \&
Puls 2008).  An important omission so far in the derivation of macroturbulence
is that time-dependent velocity phenomena also occur, besides rigid surface
rotation and turbulence. The best known example of such a phenomenon is stellar
pulsation, which causes asymmetric line-profile variations (e.g., Aerts \& De
Cat 2003).  A natural step is thus to investigate whether the needed
macroturbulence may be connected with the omission of pulsational broadening in
the line synthesis codes used for fundamental parameter estimation. In fact,
for pulsating stars along the main sequence, one also needs to add some level of
macroturbulence whenever one ignores (some of) the detected pulsations in
line-profile fitting of time-resolved or averaged spectra (e.g., Morel et al.\
2006). We investigate this hypothesis in the present paper.

\section{Computations of pulsationally broadened spectral line profiles}

\begin{figure*}
\begin{center}
\rotatebox{270}{\resizebox{13cm}{!}{\includegraphics{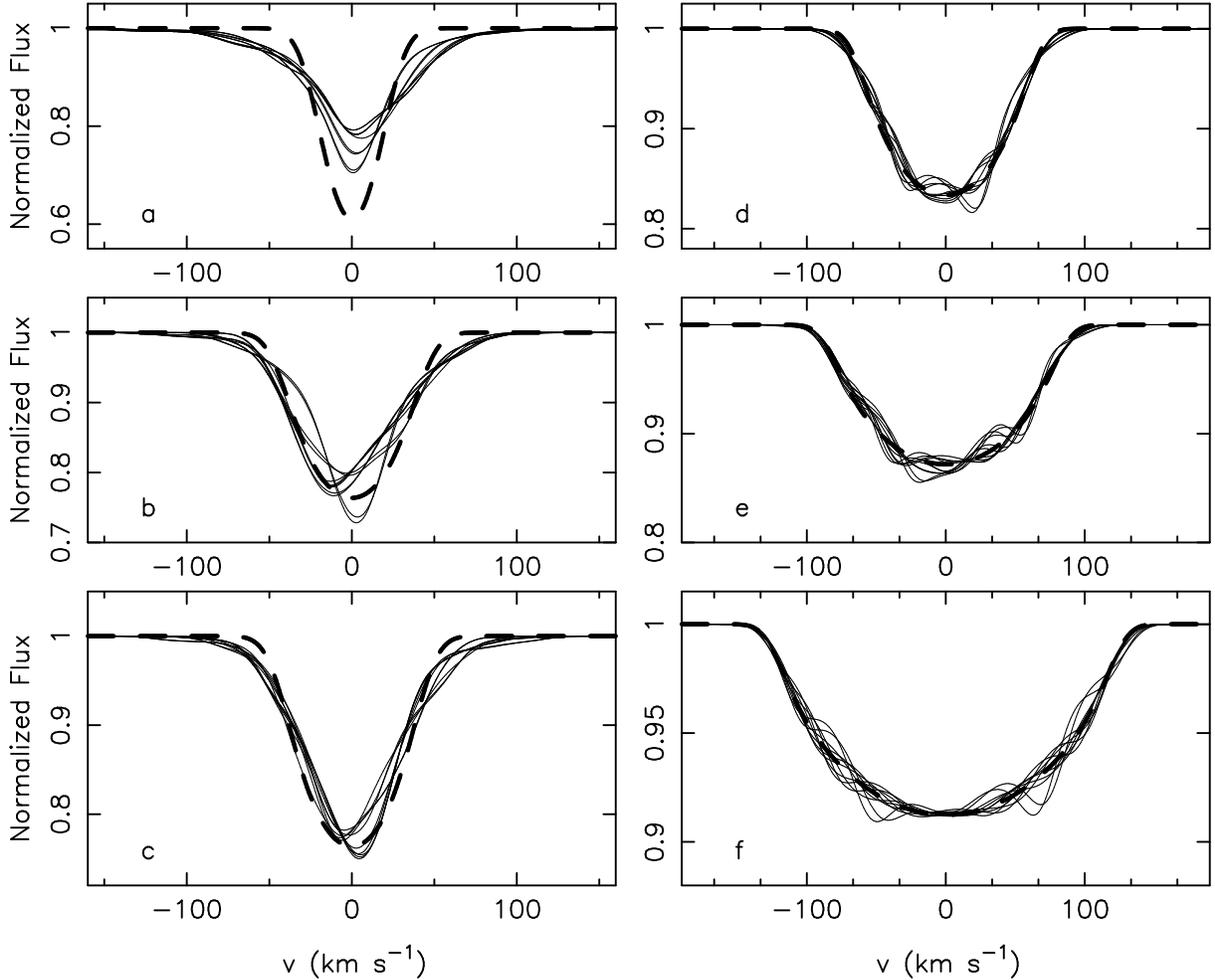}}}
\end{center}
\caption{Noiseless pulsationally and rotationally broadened profiles (thin
lines) are compared with the profile without pulsational but with rotational
broadening (dashed line).  The input parameters are the stellar inclination
angle, the amplitude of the individual modes, and the projected rotation
velocity, $(i, a, v\sin i)$.  Their values are as follows: a: $(60^\circ, 1.0,
25)$, b: $(60^\circ, 1.0, 45)$, c: $(20^\circ, 1.0, 45)$, d: $(60^\circ, 0.5,
65)$, e: $(60^\circ, 0.5, 85)$, f: $(60^\circ, 0.2, 125)$, where $a$ and $v\sin
i$ are expressed in km\,s$^{-1}$ (see text for a definition of $a$).  Note that
the line features in panels d, e, f become less visible as the S/N ratio
decreases. They essentially disappear for S/N $< 100$ (see also
Fig.\,\protect\ref{selectie}).  }
\label{profs}
\end{figure*}

Massive stars are exposed to pulsations during several phases of their life. On
or near the main sequence, these pulsations are usually driven by a heat
mechanism acting in the metal opacity bump at a temperature near some 200,000
degrees (e.g., 
Cox et al.\ 1992, Pamyatnyh 1999).  In the recent and rapidly growing
research field of asteroseismology, observed pulsations are exploited by
scientists to probe the badly known physical processes inside stars (e.g., Cunha
et al.\ 2007, Aerts et al.\ 2009), just as it was done in helioseismology for
the Sun (e.g., Gough et al.\ 1996).  Asteroseismology was proven to be a valid
tool to study the interior of massive main-sequence stars (e.g., Aerts et al.\
2003) and may imply a unique opportunity to probe the internal layers, including
the deep convection zone around the hydrogen-burning shell, of evolved stars as
well.  The discovery of gravity-mode pulsations in the B1Ib supergiant
HD\,163899 from spacebased high-precision photometry measured with the Canadian
space mission MOST (Saio et al.\ 2006) and in a sample of 40 B supergiants
(Lefever et al.\ 2007) are steps in this direction. We refer the reader to Aerts
et al.\ (2009) for a thorough description of stellar pulsation in all of its
aspects, including the particular properties of the eigenfrequencies and
eigenfunctions of pressure and gravity modes.

It is well known that stellar pulsations imply a time-dependent variation of the
shape of spectral lines (e.g., Aerts \& De Cat 2003 for a review, Chapters 4 and
6 of Aerts et al.\ 2009).  Despite this, the estimation of the rotational and
macroturbulent velocity in evolved massive stars have so far usually been done
from a single snapshot of the stellar spectrum, and assuming that no
time-dependent phenomena are present.  Here, we investigate to what extent
stellar pulsations affect the estimation of the surface rotation and
macroturbulent velocities when ignoring pulsations, as was often done so far in
the literature.  Hereto, we computed numerous sets of line profiles due to
pulsations expected in B supergiants.

\subsection{Input for the simulations}

{

Simulating line profile variations due to excited oscillations 
of a star requires the following steps:
\begin{enumerate}
\item
the computation of an equilibrium stellar structure model;
\item
the computation of the excited oscillation frequencies of the stellar model;
\item
the computation of the oscillation eigenfunctions in the line-forming region of
the stellar atmosphere;
\item
the computation of the observed line profile as seen by a distant observer,
whose line of sight is inclined with respect to the symmetry axis of the
oscillations.
\end{enumerate}

For points 1 and 2 we considered a realistic case and computed a stellar
evolution model 
}
which is representative for the evolved pulsating B1Ib star
HD\,163899 with the Code Li\'egeois d'\'Evolution Stellaire (Scuflaire et al.\
2008).  This model has the following parameters: $T_{\rm eff}=18,200$\,K,
$\log\,g=3.05$, $R/R_\odot=17.8$, $\log(L/L_\odot)=4.5$, $M/M_\odot=13$,
$Z=0.02$ and an age of thirtheen million years. It approximates well the
position of HD\,163889 in the Hertzsprung-Russell diagram (Saio et al.\ 2006).
{ 
We determined its excited pulsation modes of azimuthal order zero with a
non-adiabatic pulsation code MAD (Dupret 2001).}  
We considered all modes up to
degree ten, as it is well-known that partial geometrical cancellation effects
increase with increasing mode degree (e.g., Chapter 6 of Aerts et al.\ 2009 for
a full description of these effects) and, moreover, we needed to keep the
computation time feasible. We found 241 modes with degree $\ell$ from 1 to 10 to
be excited. All of them are gravity modes, with frequencies ranging from 0.08 to
0.68 cycles per day, a result typical for gravity modes.

{ Regarding points 3 and 4, it was shown by De Ridder et al.\ (2002) that the
temperature and gravity variations in the line-forming region due to the
pulsations do not affect the line profile variations of a non-rotating star
appreciably. This conclusion was based on the computation of temperature and
gravity variations for the stellar interior and for the atmosphere, and applying
a matching in a connecting layer which separates the region where the diffusion
approximation breaks down from the stellar interior where it is valid (Dupret et
al.\ 2002).  This justifies the use of the basic line profile theory as
described in Aerts et al.\ (1992). That framework allows the computation of the
velocity eigenvectors of the modes in the line-of-sight for a linear
limb-darkening law.  It makes use of the velocity perturbations in a single
line-forming layer of the atmosphere to predict the line profile variations,
while ignoring temperature, gravity, and rotational effects.  Ideally,
rotational effects should be included in the computations, given that the ratio
of the rotation to pulsation frequencies can be of order one.  Theories
including a non-adiabatic treatment of rotational effects due to the Coriolis
force are available for the stellar interior, where the diffusion approximation
is valid (e.g., Lee 2001, Townsend 2005).  A study as the one by Dupret et al.\
(2002) which treats the velocity and temperature perturbations of a rotating
star in the very outer atmosphere is not yet available. Developing it is beyond
the scope of the present work, which is simply to generate profiles due to
pulsations with properties similar to those observed and interpret them as
macroturbulence.  }

Rotation splits the frequencies of modes into $2\ell +1$ multiplet components
(e.g., Aerts et al.\ 2009).  As there is currently no theory to provide us with
the excitation of rotationally splitted modes, nor with the amplitudes of the
modes, we assume that all modes with azimuthal orders $m$ ranging from $-\ell$
to $\ell$ are excited with equal amplitudes 
{ 
in the line-forming region and we assume these amplitudes to be 
}
$v_{\rm p}\propto (\ell+1)^{-1}$ in
the notation by Aerts et al.\ (1992).  We checked that changing these
assumptions does not alter the conclusions presented here, by considering also
the case where only axisymmetric or sectoral modes would be excited and by
choosing different amplitude laws.  Provided that a sufficient number of modes
are included in the line broadening computations (typically at least a few
hundred), our conclusions remain the same and are thus independent of the
adopted amplitude distribution.  The conclusions are not dependent on the
particular stellar model either.

In total, the 241 excited $m=0$ modes give rise to 2965 multiplet components. We
computed the collective effect of all these 2965 gravity modes on simulated line
profiles. The simulations were made such as to mimic the effect on the
Si\,III\,4553\AA\ line in the spectrum of a star with the fundamental parameters
of HD\,163899.  In our simulations, we approximated the Si\,III\,4553\AA\ line
by a Gaussian profile of width 10\,km\,s$^{-1}$ and equivalent width of 0.25\AA,
and we adopted a linear limb darkening law with a fixed coefficient equal to
0.364 across the line. These values were also fixed when computing the fits to
the pulsationally broadened profiles. In this way, we are sure that our
conclusions on the macroturbulence are not affected by adopting a wrong
microturbulence or by a varying limb darkening coefficient across the spectral
line.  We limited ourselves to tuning towards this one Si spectral line, since
(i) it is an important diagnostical line of intermediate strength, (ii) it is
(almost) not contaminated by non-Gaussian broadening (such as Stark broadening
in the case of hydrogen and helium lines) or wind effects, and (iii) it is the
line selected for almost all of the pulsating early B stars so far as it turned
out to be best suited to derive their pulsation characteristics (Aerts \& De
Cat 2003).  On the other hand, our approximation of a constant Gaussian
intrinsic line implies that our analysis is valid for any metal line of this
width in the spectrum.  We computed time series of profiles for 50 timings taken
from a concrete line profile study ({\v S}tefl et al.\ 1999) with a total time
span of 65 days. 

\begin{figure}
\begin{center}
\rotatebox{270}{\resizebox{6.5cm}{!}{\includegraphics{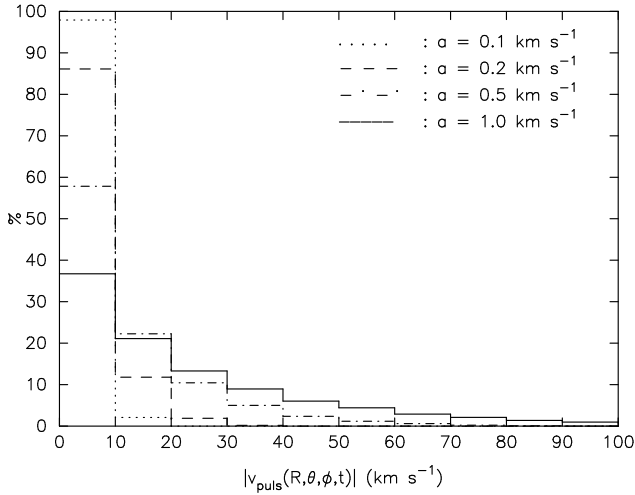}}}
\end{center}
\caption{Distribution of the projected pulsational velocity over the stellar
  surface as measured by a distant observer whose line-of-sight is inclined by
  60$^\circ$ with respect to the rotation axis of the star, for four
  distributions of the pulsational amplitudes $v_{\rm p}$ (see text for further
  explanation).  }
\label{verdeling}
\end{figure}

{We simulated various time series of 50 profiles each, taking into account
pulsational and rotational broadening, besides the intrinsic broadening of the
spectral line.  In our computations, we considered five values of the projected
rotation velocity $v\sin i$ (25,45,65,85,125\,km\,s$^{-1}$).  
{ 
We limited to this range of $v\sin i$, for which the equatorial rotation
velocities remain below 50\% of the critical velocity of the stellar model
(305\,km\,s$^{-1}$).  As can be seen in Fig.\,1 of Aerts et al.\ (2004), this
implies sufficiently small relative changes of the local radius, temperature,
gravity, and luminosity to ignore the centrifugal force in the computation of
the equilibrium structure model of the star.  }

We adopted an inclination angle $i$ between the rotational axis and the
line-of-sight of $60^\circ$, but we also considered $20^\circ$ for the case of
$v\sin i=45$\,km\,s$^{-1}$.  The symmetry axis of the pulsations was taken
equal to the rotational axis, as is usually done in pulsation studies of
non-magnetic stars. Regarding the pulsational broadening, we considered four
distributions for the intrinsic amplitude of the modes
{ in the line-forming region}: 
$v_{\rm p} =
a/(\ell+1)$ with $a=1.0,0.5,0.2,0.1$\,km\,s$^{-1}$, again using the notation of
Aerts et al.\ (1992). This means that the radial component of the pulsational
velocity vector is proportional to $v_{\rm p}$ while the transversal component
is proportional to $v_{\rm p}K$ with $K\equiv GM/\omega^2R^3$ with $G$ the
gravitational constant, $M$ and $R$ the mass and radius of the star, and
$\omega$ the angular frequency of the mode (see, e.g., Aerts et al.\ 2009). For
the adopted model we consider here, the $K-$values of the considered modes range
from 0.3 to 25.  
The choice of these amplitude distributions was made to end up
with a realistic peak-to-peak variation of the radial velocity as in published
observed time series of the few supergiant B stars for which such data are
available -- see Figs\,5 and 6 in Kaufer et al.\ (1997), Figs\,2 and 3 in Prinja
et al.\ (2004) and Fig.\,2 in Markova et al.\ (2008). These studies have led to
radial-velocity variations with peak-to-peak amplitudes between 5 and
20\,km\,s$^{-1}$. 
{
Our amplitude distributions for $v_{\rm p}$ were taken
accordingly, i.e., the collective effect of the 2965 gravity modes with the
amplitude distributions we adopted results in radial-velocity variations similar
to the observed ranges (see $\langle v\rangle$ in Fig.\,\ref{figvmacro}
discussed in Sect.\,\ref{secvmacro}).  
In this way, we are sure to have generated realistic profile variations,
irrespective of the limitations of the line profile theory discussed above.}
A summary of the input parameters of the
simulated line profile sets, along with some of their computed quantities
discussed below, is given in Table\,\ref{momenten}.

The radial velocity is an integrated quantity over the stellar surface.
Pulsating stars have time-dependent asymmetric line-profile variations.  It is
common to characterise the line profile shapes by their three lowest-order
moments, which represent the centroid velocity $\langle v\rangle$, the width
$\langle v^2\rangle$ and the skewness $\langle v^3\rangle$. A practical guide to
compute these quantities, as well as their formal definition in terms of the
surface velocity eigenfunctions, is provided in De Ridder et al.\ (2002) and
more extensively in Chapter 6 of Aerts et al.\ (2009).  Aerts et al.\ (1992) and
Aerts (1996) provided a thorough discussion of these quantities and their
suitability to interprete them in terms of pulsation theory, thus allowing an
identification of the spherical wavenumbers $(\ell,m)$ from observed time series
of moment variations.  We computed these three quantities for the simulated line
profiles, mainly to show their relation with the derived macroturbulent velocity
values obtained when ignoring the pulsational broadening, as will be discussed
in Sect.\,\ref{secvmacro}. The first moment $\langle v\rangle$ is the radial
velocity of the star, integrated over the stellar disc, with respect to the
centre of mass of the star (i.e.\ it varies around value zero during the
pulsation cycle); it is thus directly comparable to the measured radial-velocity
variations reported in the literature which are usually based on Gaussian fits
to the profiles.

In order to end up with peak-to-peak radial-velocity variations of order
20\,km\,s$^{-1}$, as measured for several supergiant B stars from metal lines,
numerous of the individual surface elements must experience a far larger
individual pulsation velocity. In the case of radial pulsations and in the
approximation of the adopted linear limb darkening law, this means that the
entire surface moves up and down with a velocity of about $20 \times 1.5 =
30$\,km\,s$^{-1}$.  This also implies that measured radial-velocity variations
above typically 40\,km\,s$^{-1}$ are the results of shock phenomena in the
atmosphere of radial B-type pulsators, leading to a so-called ``stillstand'' in
the radial-velocity curve in the case of radial modes. Examples can be found in
Aerts et al.\ (1995), Saesen et al.\ (2005) and Briquet et al.\ (2009) for the
$\beta\,$Cep stars BW\,Vul, $\xi^1$\,CMa and V1449\,Aql, respectively. Such
stillstand was so far not observed for B supergiants, so we expect the majority
of the surface elements to move subsonically (which does not imply that some
elements may encounter supersonic speeds).

For non-radial gravity modes, a wide variety of surface velocities occurs across
the stellar surface and shock phenomena are much harder to detect in integrated
quantities, such as moments or equivalent widths.  We show in
Fig.\,\ref{verdeling} the distribution of the line-of-sight components of the
total pulsational velocity vectors, which result from the addition of all the
individual vectors of the 2965 modes, for each of the surface elements (denoted
as $|v_{\rm puls}(R,\theta,\phi,t)|$), for the four amplitude sets corresponding
to the four $a$-values.  The addition of the numerous eigenvectors can result in
positive or negative projected velocity, depending on the phases of the modes
and on the place $(R,\theta,\phi)$ on the surface. We expect that, in most of
the surface points and for most of the timings, positive and negative
contributions tend to lead to a limited value of the overall pulsation velocity
due to cancelling of positive and negative mode velocities, since we assumed
there to be no phase relation between the modes.  It can be seen from
Fig.\,\ref{verdeling} that, for all four amplitude sets, the majority of the
surface elements indeed are seen to move subsonically. For
$a=1.0$\,km\,s$^{-1}$, supersonic speeds in the line-of-sight are encountered
for a considerable fraction of the surface elements, but still in less than half
of them such that the radial-velocity variations remain below 20\,km\,s$^{-1}$
(see Fig.\,\ref{figvmacro} discussed in Sect.\,\ref{secvmacro}).  The adopted
amplitude distributions thus lead to realistic peak-to-peak amplitudes for the
radial velocity.  In this way, we are sure not to overestimate the effects of
pulsations on the derivation of the macroturbulent velocity values.

We considered profile sets without noise and with white noise resulting in S/N
ratios of 200 and 500. This brings the total number of simulates profiles to
3600 (50 timings, 6 combinations $(v\sin i, i)$, 4 amplitude distributions and
3 noise levels). Examples are provided in Fig.\,\ref{profs} and lead to
the conclusion that some of the simulated profiles are considerably affected by
the collective effect of the gravity modes. In particular, the line wings are
broader than those that would occur for a star that does not have pulsations.

\subsection{Estimation of the macroturbulent velocities}
\label{secvmacro}

Various possibilities to describe macroturbulence have been presented in the
literature. We refer to Gray (2005) for a thorough discussion. In this work, we
considered an isotropic macroturbulence described by a Gaussian velocity
distribution (denoted as $A_{\rm ISO}$), as well as an anisotropic description
for which the radial and tangential velocity fields in general have a different
amplitude denoted as $A_{\rm R}$ and $A_{\rm T}$ (a so-called radial-tangential
model --- see Eq.\,(17.6), p.433 in Gray 2005). For the anisotropic model fits,
we considered the two extreme cases of allowing $A_{\rm R}$ to be free while
$A_{\rm T}=0$ and $A_{\rm R}=0$ while $A_{\rm T}$ was allowed to take any value.
In this way, each of the three models for the macroturbulence is described by
one free parameter.

For all the simulated profiles, we determined $v\sin i$ and the
macroturbulence $v_{\rm macro}$, while ignoring the presence of pulsational
broadening, as is done in the literature, by adopting a goodness-of-fit
approach.  The normalized profiles broadened by both rotation and gravity-mode
pulsations are denoted by $(\lambda_j,p_1(\lambda_j))$ and those broadened by
rotation and macroturbulence by $(\lambda_j,p_2(\lambda_j))$, with
$j=1,\ldots,N$ an index labelling the velocity pixels within the profile.  {
For the computation of $p_2$ we considered each of the three options $A_{\rm
ISO}, A_{\rm R}$, and $A_{\rm T}$.}  Each of the profiles $p_1$ and $p_2$ were
given the same equivalent width. We computed the line deviation parameter,
$\Sigma$, based on the classical statistical technique of residuals:
\begin{equation}
\displaystyle{\Sigma (v\sin i, v_{\rm macro})
\equiv \sqrt{\frac{1}{N-1} 
\sum_{j=1}^{N}
\left[p_1 (\lambda_j)-p_2 (\lambda_j)\right]^2.}}
\label{deviation}
\end{equation}
This quantity is the standard deviation of the residual profile $|p_1-p_2|$,
averaged over all velocity pixels in the line profile, expressed in continuum
units. It is thus a measure of the fit quality, directly interpretable in terms
of the S/N ratio of measurements.  The optimal choice of the parameters 
$(v\sin i, v_{\rm macro})$ 
is then found by carefully screening the 2-dimensional
parameter space in $v\sin i$ and $v_{\rm macro}$ (in steps of
1\,km\,s$^{-1}$ for each of $v\sin i$ and $v_{\rm macro}$) and by
identifying $\Sigma_{\rm m}\equiv\min_{(v\sin i, v_{\rm macro})} \Sigma$,
{ where $v_{\rm macro}$ can be any of $A_{\rm ISO}, A_{\rm R}$ or $A_{\rm
T}$. Moreover, we allowed two options to fit the pulsationally broadened profile
$p_1$: the one such that the wavelength position of the minimum 
of $p_1$ and $p_2$ coincide and the one
such that their first moments $\langle v\rangle$ are in best agreement.  Note
that these two options are equivalent only in the case of symmetric
profiles. For each generated profile, we performed these six fitting exercises,
keeping in each case the fit with the lowest $\Sigma_{\rm m} (v\sin i,
v_{\rm macro})$ as the best one.  We then compared the parameters $(v\sin i,
v_{\rm macro})$ of the best fitting profile $p_2$ with the input $v\sin i$
of $p_1$ and analysed the values of $v_{\rm macro}$. Six prototypical examples
of best fits are shown in Fig.\,\ref{selectie}.
}
\begin{figure*}[t!]
\begin{center}
\rotatebox{270}{\resizebox{13cm}{!}{\includegraphics{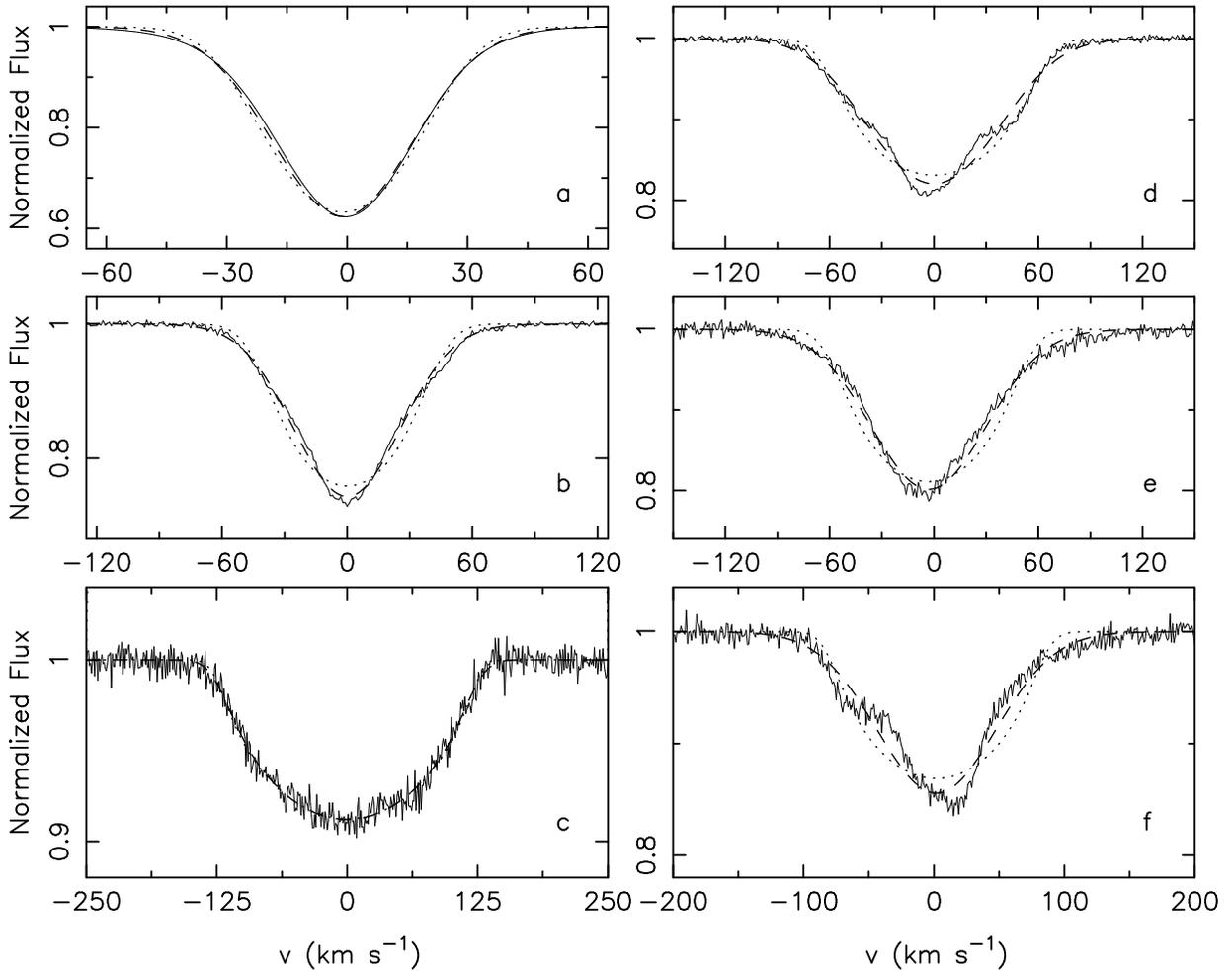}}}
\end{center}
\caption{Six pulsationally broadened profiles with different S/N ratio (full
lines) are compared with their best fit including both rotation and 
macroturbulence (dashed lines) and rotation alone (dotted line). The values for
the input rotation velocity, the rotation velocity from a fit without
macroturbulence, and from a fit with isotropic macroturbulence, 
$(v\sin i({\rm in}), v\sin i({\rm fit}, v_{\rm macro}=0); 
v\sin i({\rm fit}), 
v_{\rm macro})$, are as follows: 
a: $(25, 27; 8, 14)$,
b: $(45, 44; 11, 23)$,
c: $(125, 126; 125, 10)$,
d: $(65, 64; 49, 24)$,
e: $(45, 57; 5, 32)$,
f: $(85, 82; 14, 44)$,
where all velocities are expressed in km\,s$^{-1}$. 
{ The fits without macroturbulence (dotted lines) lead to more reliable
values of $v\sin i$ (see text for explanation). }
}
\label{selectie}
\end{figure*}

\begin{figure*}
\begin{center}
\rotatebox{270}{\resizebox{13cm}{!}{\includegraphics{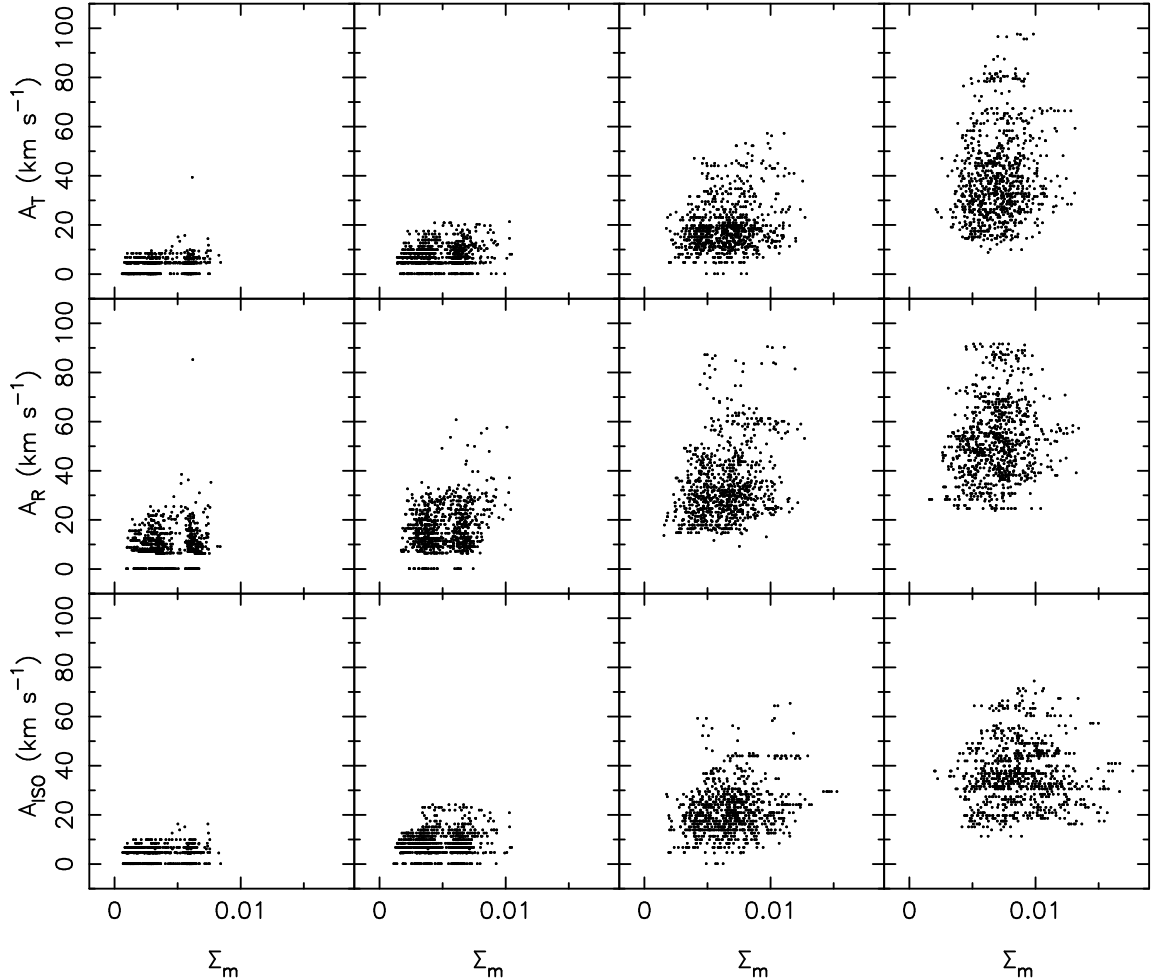}}}
\end{center}
\caption{
$\Sigma_{\rm m}$ (in continuum units) for the three models of macroturbulence
$A_{\rm ISO},  A_{\rm R}$, and $A_{\rm T}$, for the four amplitude distributions
(left to right: $a=0.1, 0.2, 0.5, 1.0\,$km\,s$^{-1}$, see also
  Table\,\protect\ref{momenten}).  
}
\label{sigmam}
\end{figure*}

\begin{figure}[h!]
\begin{center}
\rotatebox{270}{\resizebox{6.5cm}{!}{\includegraphics{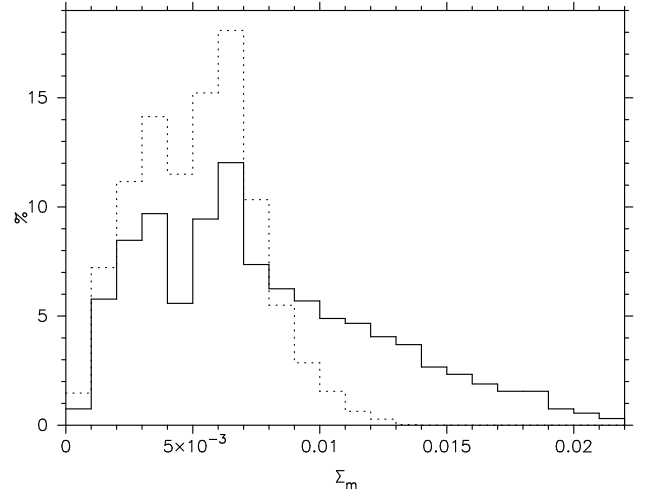}}}
\end{center}
\caption{ Distribution of $\Sigma_{\rm m}$, which is a measure of the fit
quality averaged over the line profile and expressed in continuum units -- see 
Eq.\,(\ref{deviation}),
 for 3600 fits 
without macroturbulence
(full line) and with macroturbulence (either $A_{\rm ISO}, A_{\rm
R}$, or $A_{\rm T}$ whichever gave the lowest $\Sigma_{\rm m}$, dotted
line).   }
\label{sigmam2}
\end{figure}

\begin{figure*}
\begin{center}
\rotatebox{270}{\resizebox{13cm}{!}{\includegraphics{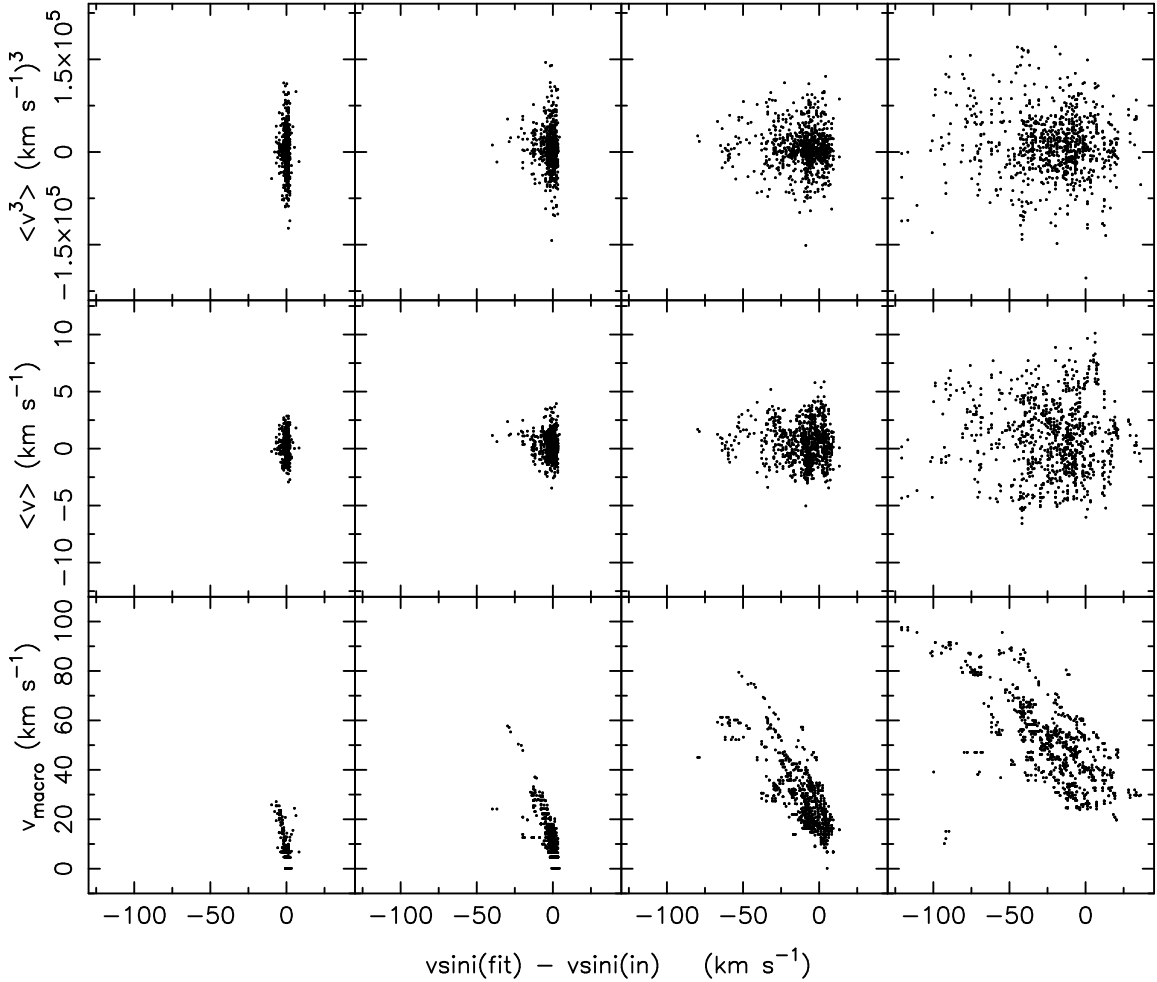}}}
\end{center}
\caption{ The macroturbulence (either $A_{\rm ISO}, A_{\rm R}$, or $A_{\rm T}$)
resulting in the minimal $\Sigma_{\rm m}$, as well as the first and third
moments, as a function of the excess of the rotational velocity estimate
compared with the input value, derived from line profile fits ignoring the
presence of pulsational broadening, for the four amplitude distributions (left
to right: $a=0.1, 0.2, 0.5, 1.0\,$km\,s$^{-1}$, see also
Table\,\protect\ref{momenten}). For profiles unaffected by pulsational
broadening, each of the macroturbulence, $\langle v\rangle$, and $\langle
v^3\rangle$ is zero. }
\label{figvmacro}
\end{figure*}

\begin{table*}
\caption{\label{momenten} Ranges of the intervals for the macroturbulence and
  moments of the 72 sets of simulated line profiles with pulsational broadening
  (3 values of the S/N for each of the combinations ($i$, $v\sin i$, $a$), with
  $v_{\rm p} = a/(\ell+1)$. See text for further explanation.}
\begin{center}
\begin{tabular}{ccccccc}
\hline
$i$ & $v\sin i$ &  $a$ & $v_{\rm macro}$ & $\langle v\rangle$ &  $\langle v^2\rangle$ & $\langle v^3\rangle$\\
$^\circ$ & km\,s$^{-1}$ & km\,s$^{-1}$ &  km\,s$^{-1}$ & km\,s$^{-1}$ &   km$^2$\,s$^{-2}$ &  km$^3$\,s$^{-3}$ \\
\hline
20 & 45 & 0.1 & [0,14] & [-0.5,0.7] &  [588,611] & [-920,1645] \\
20 & 45 & 0.2 & [0,21] & [-1.0,1.4] &  [618,674] & [-1826,3287]\\
20 & 45 & 0.5 & [7,44] & [-2.4,3.6] &  [788,1091] & [-7956,12931]\\
20 & 45 & 1.0 & [21,67] & [-4.8,7.1] & [1360,2543] & [-32298,64385]\\
\hline
60 & 25 & 0.1 & [0,7] & [-0.3,0.4] & [248,272] & [-558,544] \\
60 & 25 & 0.2 & [0,12]& [-0.7,0.7] & [255,328] & [-804,1586] \\
60 & 25 & 0.5 & [11,29] & [-1.6,1.8] & [326,691] & [-3872,10999] \\
60 & 25 & 1.0 & [24,56] & [-3.3,3.6] & [573,1942] & [-16257,59011] \\
\hline
60 & 45 & 0.1 & [0,8] & [-0.5,0.6] & [576,602] & [-971,1081] \\
60 & 45 & 0.2 & [0,17] & [-0.9,1.2] & [591,656] & [-1987,2804] \\
60 & 45 & 0.5 & [11,33] & [-2.3,2.9] & [688,1068] & [-8721,9546] \\
60 & 45 & 1.0 & [26,67] & [-4.7,5.9] & [982,2599] & [-41304,29635] \\
\hline
60 & 65 & 0.1 & [0,23] & [-0.3,0.7] & [1095,1143] & [-1725,3343] \\
60 & 65 & 0.2 & [0,28] & [-0.7,1.5] & [1106,1225] & [-3999,6820] \\
60 & 65 & 0.5 & [6,53] & [-1.7,3.6] & [1176,1672] & [-12753,19100] \\
60 & 65 & 1.0 & [25,70]& [-3.3,7.3] & [1466,3266] & [-42861,56467] \\
\hline
60 & 85 & 0.1 & [0,14] & [-0.5,0.8] & [1793,1848] & [-3349,6224] \\
60 & 85 & 0.2 & [0,33] & [-1.0,1.7] & [1794,1912] & [-7028,11406] \\
60 & 85 & 0.5 & [12,61] &[-2.4,4.1] & [1853,2314] & [-20976,27259] \\
60 & 85 & 1.0 & [27,81] &[-4.7,8.3] & [2173,3772] & [-61905,93998] \\
\hline
60 & 125 & 0.1 & [0,22] & [-0.6,0.7] & [3745,3865] & [-8440,11892] \\
60 & 125 & 0.2 & [0,55] & [-1.1,1.4] & [3707,3956] & [-16857,23578] \\
60 & 125 & 0.5 & [14,79] & [-2.7,3.4] & [3665,4422] & [-44830,64179] \\
60 & 125 & 1.0 & [10,97] & [-5.5,6.7] & [3810,6007] & [-124040,166820] \\
\hline
\end{tabular}
\end{center}
\end{table*}

In Fig.\,\ref{sigmam} we show the outcome of the fit to the 3600 simulated
profiles, for the three models we considered for the macroturbulence.  For all
three models, it was found that the inclusion of an ad-hoc macroturbulence
parameter leads to better fits than those obtained when only allowing
rotational broadening, which has to be the case given that there is one more
degree of freedom. This is visible from Fig.\,\ref{sigmam2} where we show the
{
distribution of $\Sigma_{\rm m}$ deduced from fits with and without allowing a
parameter for macroturbulence. Figure\,\ref{sigmam2} contains all simulated
profiles; these two global distributions are the same as those for the five
separate values of $v\sin i$, which is as expected given that the pulsational
broadening was simply added to the rotational broadening without any coupling
between the two.  One would typically improve the fit quality from eye
inspection by incorporating macroturbulence for $\Sigma_{\rm m}>0.008$; this
corresponds with the dotted lines in panels b,d,e,f in Fig.\,\ref{selectie}
whose counterpart with macroturbulence represented by the dashed lines imply a
noticable reduction in $\Sigma_{\rm m}$ (for comparison, 
the dotted lines in panels a and c have
$\Sigma_{\rm m}=0.0058$ and 0.0056, respectively, and would probably not give
rise to the introduction of macroturbulence).
It can be seen from the distribution of
$\Sigma_{\rm m}$ in Fig.\,\ref{sigmam2} that {\it the fit quality is very good
for the large majority of profiles when allowing for macroturbulent broadening.}
In 89\% of the cases, the fit with macroturbulence has $\Sigma_{\rm
m}<0.008$. If we do not include macroturbulence, 59\% of the fits has
$\Sigma_{\rm m}<0.008$.  }

Returning back to Fig.\,\ref{sigmam}, we deduce that the lowest $\Sigma_{\rm m}$
values were reached for $A_{\rm ISO}, A_{\rm R}$, and $A_{\rm T}$ in 1224, 1343,
and 1033 of the cases, so these descriptions are basically equivalent in
appropriateness to mimic pulsational broadening.  It can be seen from
Fig.\,\ref{sigmam} that the radial model $A_{\rm R}$ needs higher values to
achieve a good fit compared with the isotropic and tangential model. This is
logical, because the pulsation velocities of the modes are dominantly horizontal
in nature.  There are some differences between the fit quality in a global sense
for three considered models, but the main conclusion is that {\it the missing
broadening caused by the pulsations can often only be compensated by quite large
values of the macroturbulence}.

In Fig.\,\ref{figvmacro} we show the value of the macroturbulence with the
lowest $\Sigma_{\rm m}$, along with the first and third velocity moment, of the
simulated profiles. As Fig.\,\ref{sigmam}, this plot illustrates that the value
of the macroturbulence can be very large, compatible with what is found in the
literature, if one ignores pulsational broadening, even though the centroid
velocity variations $\langle v\rangle$ induced by the pulsations are modest. The
reason for this is that the line width is a function of the square of the
velocity, and the line skewness is represented by $\langle v^3\rangle$.  Thus,
one needs to compensate the line width and line wing shape by a large value for
the macroturbulence whenever one wants to achieve a good profile fit.

The reported absence of line asymmetries in the literature must be compatible
with our physical line-broadening model. First of all, most spectroscopic
studies in which symmetric profiles are mentioned rely on visual inspection of
only one spectrum, while line profile variability (and thus asymmetry) is almost
always found when multiple-epoch observations are taken.  Visual inspection of
the fits in the top and bottom of the right column in Fig.\,\ref{selectie}
reveals line asymmetry from one snapshot spectrum, while the other four profiles
might give the impression of being symmetric. Typically, the profiles simulated
with $a=1.0$\,km\,s$^{-1}$ would be detectable by visual inspection of the
profiles.  However, when one computes diagnostic line quantities, it often
becomes obvious that the lines deviate from symmetry even if seemingly symmetric
by eye. The best diagnostic parameters to characterise line asymmetry in the
case of pulsations are the line moments.  While line bisectors and velocity
spans are often used in the cool star and exoplanet communities, such parameters
are not suited to be interpreted in terms of pulsational parameters while
moments are (e.g., Aerts et al.\ 1992, Dall et al.\ 2006, Hekker et al.\
2006). In practice one can use the property that the odd moments of a
symmetrical line profile are zero. Thus, the values of $\langle v\rangle$ and
$\langle v^3\rangle$ of metal lines measured with a high resolution and high S/N
ratio are well suited to decide if an observed line profile is subject to
time-dependent line asymmetry whenever this is not obvious from visual
inspection.  As a guide, we provide the ranges of the values of the moments of
the generated profile sets in Table\,\ref{momenten}.  The values of $\langle
v\rangle$ for the six profiles shown as full lines in Fig.\,\ref{selectie} are
a: $-1.1$, b: 0.8, c: $-0.3$, d: $-1.0$, e: $-1.4$, and f: 1.9\,km\,s$^{-1}$.
The corresponding values of $\langle v^3\rangle$ are a: $-1805$, b: 226, c:
9964, d: $-45658$, e: $-36972$, and f: 12470\,km$^3$\,s$^{-3}$. All these values
would be zero in the case of symmetrical profiles subject to white noise.  The
deviation is small, of order a few km\,s$^{-1}$, for $\langle v\rangle$, because
this quantity measures the centroid of the line and thus is
independent of $v\sin i$ and the microturbulence, while these
two quantities do affect $\langle v^3\rangle$ (see Aerts et al.\ 1992).

The use of the odd moments for asymmetry detection has to be done
from a few spectra spread at least a few days in time, because line blending of
course also causes a deviation from symmetry. Such deviation is
time-independent, though, while the signature of pulsational broadening is
always time-dependent and has typical periodicities of several hours to 
a few days in hot massive
stars.

\begin{figure*}
\begin{center}
\rotatebox{270}{\resizebox{5.5cm}{!}{\includegraphics{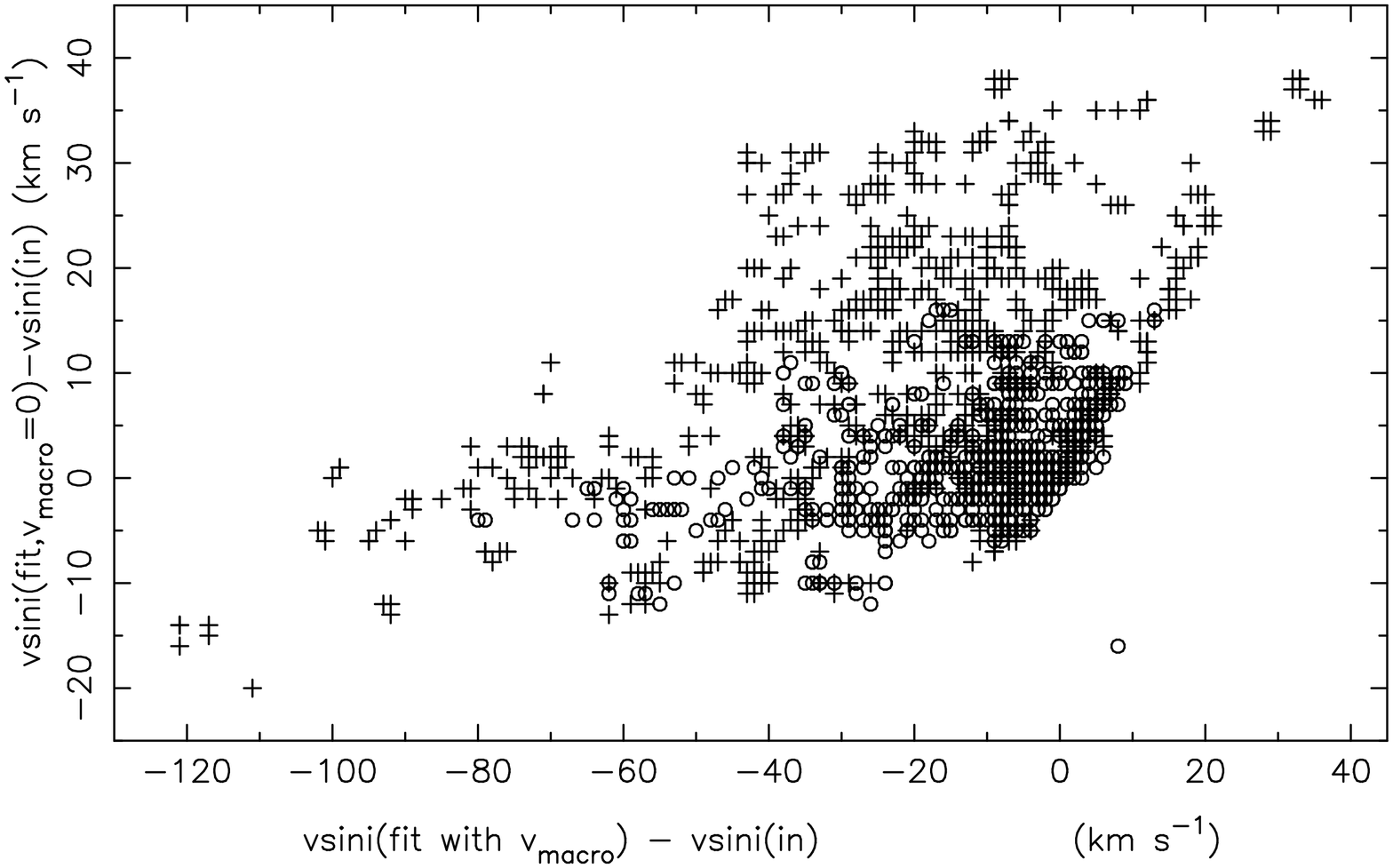}}}
\rotatebox{270}{\resizebox{5.5cm}{!}{\includegraphics{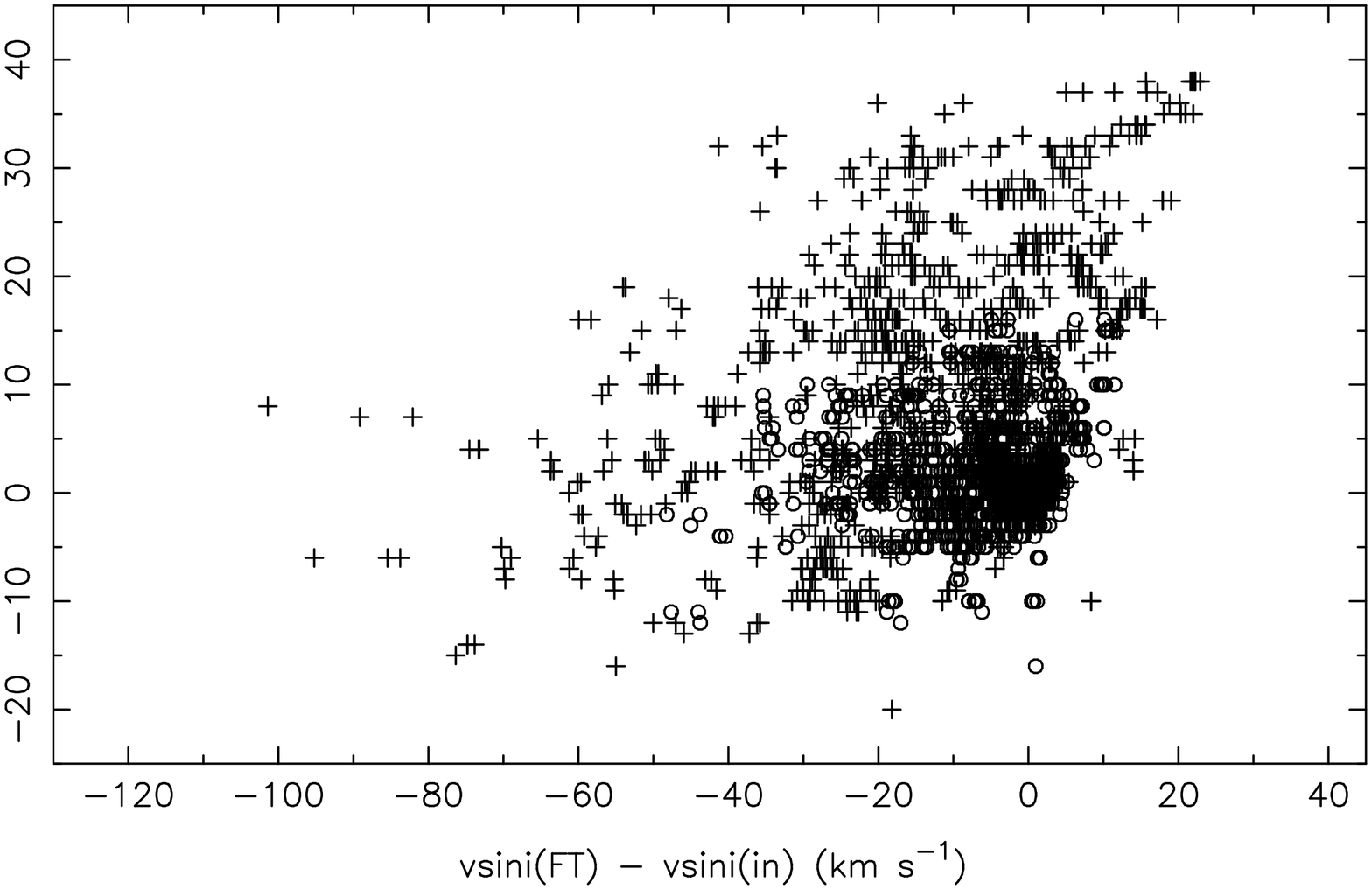}}}
\end{center}
\caption{Left: the estimated minus input rotation velocity from a fit without
macroturbulence as a function of a fit with macroturbulence, for the simulations
described in the text. The two 
different symbols indicate simulations for the four
amplitude distributions $a=0.1, 0.2, 0.5 (\circ)$, $1.0
(+)$\,km\,s$^{-1}$ as explained in the text and listed in
Table\,\protect\ref{momenten}. Right: the estimated minus input rotation
velocity from a fit without macroturbulence as a function of the value derived
from the Fourier Transform method.  }
\label{vsini}
\end{figure*}

\begin{figure*}
\begin{center}
\rotatebox{270}{\resizebox{6.5cm}{!}{\includegraphics{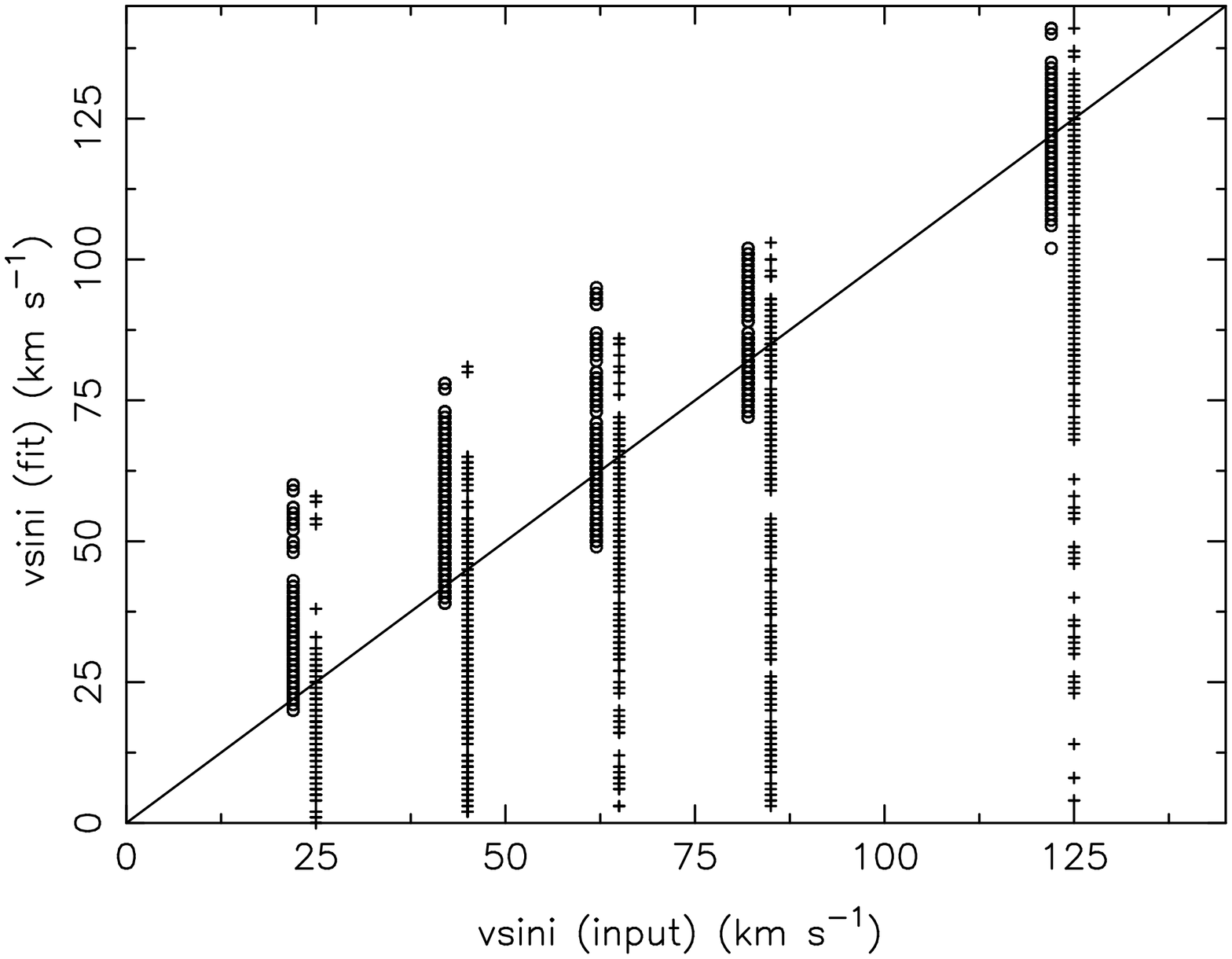}}}
\rotatebox{270}{\resizebox{6.5cm}{!}{\includegraphics{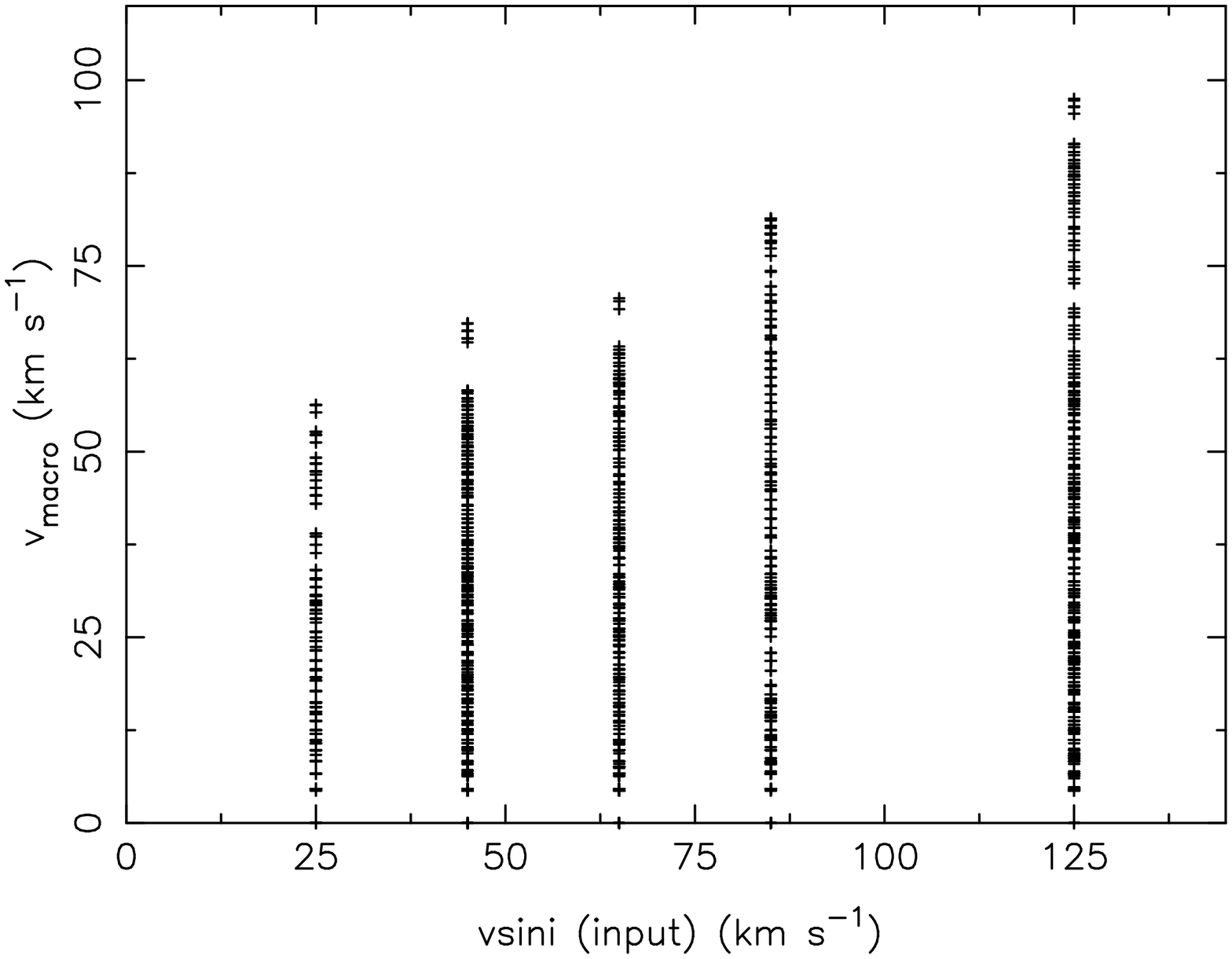}}}
\end{center}
\caption{
Left: $v\sin i$ for fits without
macroturbulence (circles) and with macroturbulence (plus signs) as a function of
the input $v\sin i$, for all the simulated line profiles.  The circles were
artificially shifted to slightly lower input $v\sin i$ for visibility purposes.
Right: the corresponding values for $v_{\rm macro}$.}
\label{inputvsini}
\end{figure*}

\subsection{Consequences for the rotational velocity estimate}
\label{secvsini}

\begin{figure*}
\begin{center}
\rotatebox{270}{\resizebox{13cm}{!}{\includegraphics{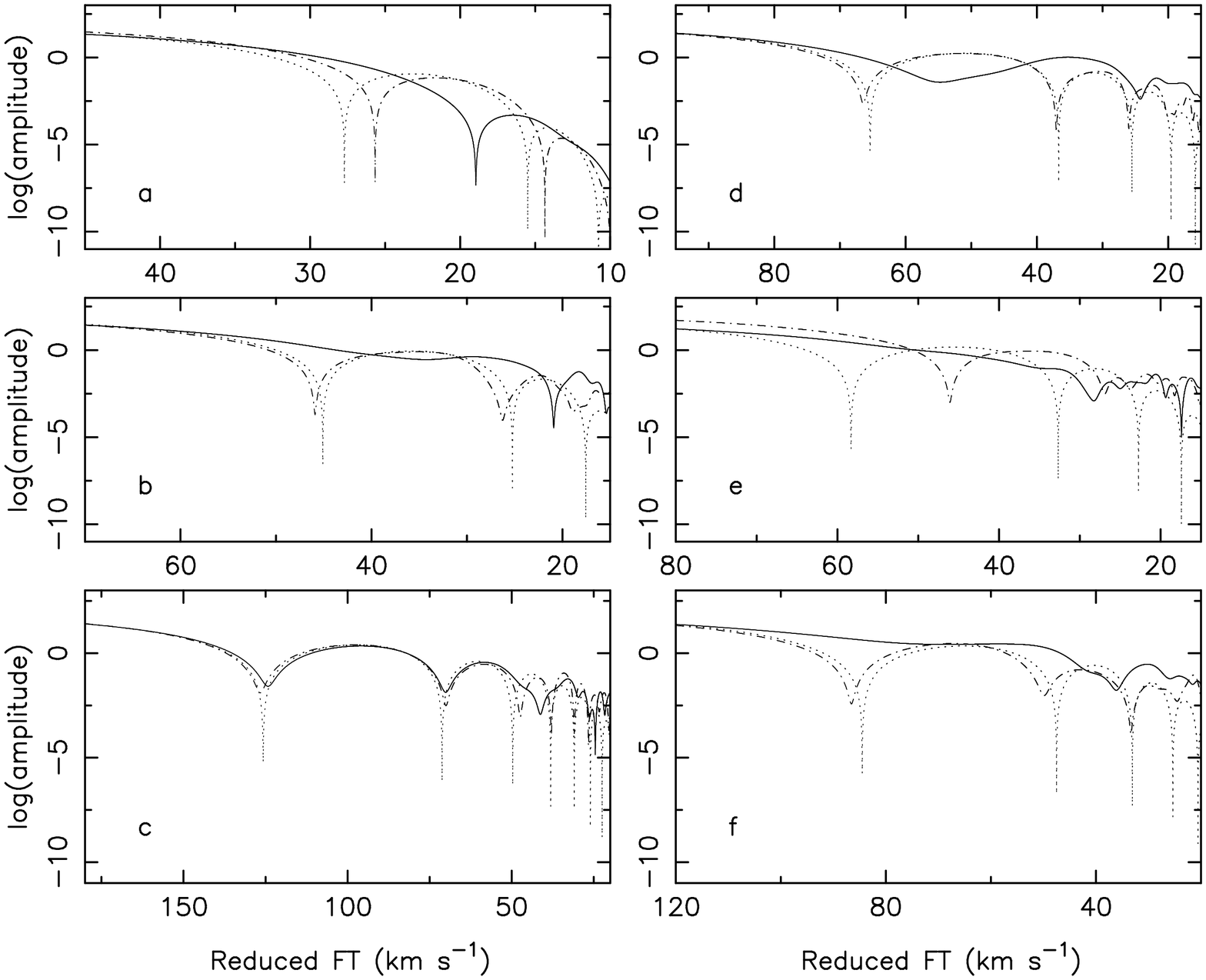}}}
\end{center}
\caption{Fourier transforms reduced to velocity units for the profiles in
Fig.\,\protect\ref{selectie}.  The full and dotted lines have the same meaning
as in Fig.\,\protect\ref{selectie}. The dashed-dotted lines represent the
results of a profile with the input rotational velocity and the same S/N ratio
as the full lines. By using the first (i.e., leftmost) minimum of the Fourier
transform to derive $v\sin i$, profiles affected by pulsational broadening would
mostly result in too low values compared to the actual (input) ones
(corresponding to the first minima of the dashed-dotted lines). On the other
hand, $v\sin i$ values derived from profile fits assuming $v_{\rm macro}=0$
(dotted) often compare quite well to the actual values.  }
\label{selectieft}
\end{figure*}

An important conclusion based on Fig.\,\ref{figvmacro}
is that the inclusion of macroturbulence to obtain a line fit may
result in a serious underestimation of the true projected rotational velocity,
irrespective of which description for $v_{\rm macro}$ is used.
We encountered mismatches compared with the input $v\sin i$ above
100\,km\,s$^{-1}$. The question thus arrises if it is not wiser to exclude
a $v_{\rm macro}$ parameter to search the best value of $v\sin i$ or to resort
to other methods to determine this parameter.

The rotation velocities derived from a fit to the pulsationally broadened
profiles with the inclusion of macroturbulence (we selected the version of
$A_{\rm ISO}$, $A_{\rm R}$, and $A_{\rm T}$ which led to the lowest $\Sigma_{\rm
m}$ for the plot) and without it are compared with the input $v\sin i$ in the
left panel of Fig.\,\ref{vsini}.  It can be seen that the mismatch of the
rotation velocity ranges from -20 to 40\,km\,s$^{-1}$ for fits without allowing
a parameter for macroturbulent broadening while it ranges from $-120$ to
40\,km\,s$^{-1}$ if a parameter for macroturbulence is allowed. From this we
conclude {\it it is better to avoid the inclusion of macroturbulence in a
goodness-of-fit approach as in Eq.\,(\ref{deviation}) when the goal is to
achieve a good estimate of $v\sin i$}. Even in that case, $v\sin i$ may be quite
wrong when derived from profiles which are pulsationally broadened.

In view of the importance of having an appropriate $v\sin i$ estimate, we also
resorted to the popular Fourier Transform (FT) method. This method was
introduced by Gray (1973, 1975). It was evaluated specifically for hot massive
stars by Sim{\'o}n-D{\'{\i}}az \& Herrero (2007).  It allows to estimate $v\sin
i$ from the first minimum of the FT of a line profile.  What is often forgotten,
however, is that its basic {\it assumption\/} is that the line profiles are {\it
symmetric}, which is not the case when pulsations (or other phenomena like
spots) occur (e.g., Smith \& Gray 1976). We thus investigated how robust the
method is when this condition is not met. We applied the method to all the 3600
pulsationally broadened profiles and derived $v\sin i$ by careful visual
inspection of their FTs. A few of the FTs are shown in Fig.\,\ref{selectieft}
while the global mismatch in $v\sin i$ is compared with the one obtained from
the goodness-of-fit method for $v_{\rm macro}=0$ in the right panel of
Fig.\,\ref{vsini}.  We see that the FT method outperformes the goodness-of-fit
method when $v_{\rm macro}$ is allowed for. On the other hand, the ability of
the FT method to estimate $v\sin i$ is also affected by the pulsational
broadening for a fraction of the simulated profiles and also leads to too low
estimates for $v\sin i$.  The offsets between the input value of $v\sin i$ and
its estimate from the goodness-of-fit method with $v_{\rm macro}=0$ on the one
hand, and from the FT method on the other hand, are above 10\,km\,s$^{-1}$ in
16\% and 26\% of the 3600 cases, respectively.  For mismatches above
20\,km\,s$^{-1}$ these numbers decrease to 6\% and 12\%, and above
30\,km\,s$^{-1}$ a further decrease to 1.7\% and 5\% occurs.

As illustrated in Fig.\,\ref{selectieft} for a few cases, the results of the FT
method improve appreciably when applied to the best fit of the line profile
including only microturbulent and rotational broadening, i.e., without allowing
macroturbulence. The reason is that, in this case, one approximates the true
pulsationally broadened asymmetric profile by one which is symmetric and has
less extended wings such that the basic assumption of the FT method is
fulfilled.  It was already emphasized long ago by Mihalas (1979) that the FT
method has limitations of applicability when various broadening functions are
convolved and result in skew profiles, which is the situation we encounter here
for the gravity modes. The FT method is reliable in filtering out the value of
$v\sin i$ from the observed spectral lines, when the rotational broadening is
very dominant while the pulsational amplitudes are very low 
(as in panel c of Fig.\,\ref{selectieft}) or when broadening due
to spots or pressure modes occurs, which leave the line wings almost unaltered
and affect mainly the central parts of the lines.

In Fig.\,\ref{inputvsini} we compare the input value of $v\sin i$ with the value
deduced from a fit with and without allowing a parameter for macroturbulence; we
also show the corresponding values of $v_{\rm macro}$ for each of the simulated
profiles.  We find an overestimation of the rotation from $\Sigma_{\rm m}$ for
low input $v\sin i$, because we need to compensate for the pulsational
broadening and this can only be achieved by fitting a profile with a higher
$v\sin i$ than the input value. When allowing for a macroturbulence, however, we
cover the entire range of projected rotation velocities between zero and values
up to some 20\,km\,s$^{-1}$ above the input value of $v\sin i$, i.e., for
several cases a serious underestimation of the true rotational velocity
occurs. This mismatch increases with the input $v\sin i$ and occurs whenever the
wings of the profiles are severaly broadened due to a positive interference of
the modes with the largest horizontal velocity amplitude at some timings in the
beat cycle and/or when large line asymmetries occur (see, e.g., panels d, e, f
of Fig.\,\ref{selectie} which typically have large values of $|\langle
v^3\rangle |$). The right panel of Fig.\,\ref{inputvsini} shows that large
values of $v_{\rm macro}$ occur for all input values of $v\sin i$, but the most
extreme values for $v_{\rm macro}$ occur typically for the broader profiles due
to rotation and the larger pulsational amplitudes (see also Fig.\,\ref{sigmam}).

We come to the important conclusion that, in the case of line profile broadening
due to gravity modes, $v\sin i$ estimates are best derived
from a simple goodness-of-fit to observed profiles, including only
microturbulence and rotational broadening and no macroturbulence.
}

\section{Implications}

The idea that macroturbulence originates from stellar pulsations is not new.  In
fact, Lucy (1976) already suggested pulsations as a potential possibility to
explain macroturbulence. Unfortunately, he did not have the observational
capabilities nor the theoretical development to study the effects of pulsations
on line profiles.  Recent observations of massive stars with the CoRoT space
mission indeed reveal the occurrence of hundreds of pulsation modes with
white-light amplitudes in the range 0.01 and 0.1\,mmag which went unnoticed in
ground-based data (Degroote et al.\ 2009a,b).  Moreover, the discovery of
massive pulsators in low-metallicity environments (e.g., Ko{\l}aczkowski et al.\
2006, Narwid et al.\ 2006, Sarro et al.\ 2009) also shows that current
excitation computations (Miglio et al.\ 2007) still underestimate the number of
excited modes. Our results are thus also relevant for evolved stars in the
Magellanic Clouds.

As an important side result of our study, we conclude that the rotational
velocities of evolved massive stars can be seriously underestimated by using
line profile fits based on a description in terms of
macroturbulence. Ironically, this finding is just opposite to previous arguments
that, by neglecting macroturbulence, the derived $v\sin i$ values are
significantly overestimated. In order to avoid erroneous estimates of $v\sin i$,
we advise to compute the moments of the line profiles as well as to compare the
values of $v\sin i$ from fits with and without allowing macroturbulent
broadening both by a goodness-of-fit approach and by the Fourier method. In this
way, the probability of a wrong $v\sin i$ estimate is relatively low.

It is remarkable that the link between pulsational broadening and
macroturbulence, and its effect on the derivation of $v\sin i$, was never
thoroughly investigated, particularly since the surface rotational velocity
derived from line profile fitting constitutes a crucial stellar parameter which
is used to evaluate stellar evolution theory.  Several authors, among which
Hunter et al.\ (2008), claim to have found too low observed rotational
velocities for evolved massive stars compared with theoretical predictions.  Our
physical model of collective pulsational broadening may help resolve this
discrepancy.  Accurate derivations of the rotational velocity of massive stars
are also relevant in the context of Gamma-Ray-Burst progenitor studies (e.g.,
Yoon et al.\ 2006). In view of our results, we strongly advise to use
multi-epoch observations, because that is the best way to estimate the effect of
pulsational broadening.  One should attempt to take at least ten spectra with a
resolution above 30,000 and a S/N ratio above 200, spread over different nights,
and consider the broadening of different metal lines, to achieve a valid
estimate of the surface rotation.

{ Our present study was based on simulations in which we considered
pulsational line broadening due to velocity perturbations ignoring the Coriolis
force, which were then interpreted as macroturbulence.  The resulting simulated
profiles were constructed in such a way as to lead to realistic radial-velocity
variations.  It might be worth to investigate how the inclusion of the
collective effect of non-adiabatic temperature and gravity variations in the
line-forming region of a star subject to the Coriolis force will affect the line
wing broadening and its interpretation in terms of macroturbulence. Irrespective
of the limitations of present line profile theory, our conclusion is clear:
ignoring time-dependent pulsational line broadening in line profile fits of
snapshot spectra may lead to the need to introduce an ad-hoc velocity field to
account for the missing broadening in the line wings.  This implies the risk of
a wrong estimation of the projected rotational velocity of the star.}

\begin{acknowledgements}
The research leading to these results has received funding from the European
Research Council under the European Community's Seventh Framework Programme
(FP7/2007--2013)/ERC grant agreement n$^\circ$227224 (PROSPERITY), as well as
from the Research Council of K.U.Leuven grant agreement GOA/2008/04. 
The computations for this research have been done on the VIC HPC supercomputer
of the K.U.Leuven. CA is much indebted to Dr.\ Leen Decin for explaining her how
to use the VIC and to Dr.~Karolien Lefever for valuable discussions. We 
acknowledge suggestions from the referee which improved our paper.
\end{acknowledgements}

\end{document}